\documentclass[fleqn,usenatbib]{mnras}

\usepackage{newtxtext,newtxmath}

\usepackage[T1]{fontenc}

\DeclareRobustCommand{\VAN}[3]{#2}
\let\VANthebibliography\thebibliography
\def\thebibliography{\DeclareRobustCommand{\VAN}[3]{##3}\VANthebibliography}

\usepackage{graphicx}	
\usepackage{amsmath}	
\usepackage{color}
\usepackage{xcolor}

\definecolor{amethyst}{rgb}{0.6, 0.4, 0.8}
\definecolor{orange}{rgb}{0.89, 0.26, 0.2}

\title[BCGs velocity dispersion in simulations]{Velocity dispersion of brightest cluster galaxies in cosmological simulations}

\author[I. Marini et al.]{
I. Marini,$^{1,2,3,4}$\thanks{E-mail: ilaria.marini@inaf.it} S. Borgani,$^{1,2,3,4}$ A. Saro,$^{1,2,3,4}$ G. L. Granato,$^{2,5,3}$ C. Ragone-Figueroa,$^{5,2}$  B. Sartoris,$^{2,3}$ \\~\\
{\rm {\LARGE K. Dolag, $^{6}$ G. Murante,$^{2,3}$ A. Ragagnin,$^{7,2,3}$ Y. Wang$^{8}$}}
\\
    $^1$ Astronomy Unit, Department of Physics, University of Trieste, via Tiepolo 11, I-34131 Trieste, Italy\\
    $^2$ INAF-Osservatorio Astronomico di Trieste, via G. B. Tiepolo 11, I-34143 Trieste, Italy\\
    $^3$ IFPU - Institute for Fundamental Physics of the Universe, Via Beirut 2, 34014 Trieste, Italy\\  
    $^4$ INFN--Sezione di Trieste, Trieste,  Italy\\
    $^5$ Instituto de Astronom\'ia Te\'orica y Experimental (IATE), Consejo Nacional de Investigaciones Cient\'ificas y T\'ecnicas de la\\
Rep\'ublica Argentina (CONICET), Universidad Nacional de C\'ordoba, Laprida 854, X5000BGR, C\'ordoba, Argentina\\
    $^6$ Universitäts-Sternwarte München, Fakultät für Physik, LMU Munich, Scheinerstr. 1, 81679 München, Germany\\
    $^7$ Dipartimento di Fisica e Astronomia "Augusto Righi", Alma Mater Studiorum Università di Bologna, via Gobetti 93/2, I-40129 Bologna, Italy\\
    $^8$ School of Physics and Astronomy, Sun Yat-sen University, 519082, Zhuhai, Guangdong Province, China\\
}

\date{Accepted 2021 August 28. Received 2021 August 18; in original form 2021 June 28}

\pubyear{2021}

\def \dispBCG{{$\sigma_\mathrm{BCG}^{\star}$}}
\def \dispCL{{$\sigma_{200}$}}

\def \m200{{$M_{200}$}}
\def \r200{{$R_{200}$}}
\def \ll'{{``}}
\def \rr'{{''}}
\def \th{{$^{th}$}}
\def \msun{{$M_{\odot}$}}

\usepackage{graphicx}

\begin{document}
\label{firstpage}
\pagerange{\pageref{firstpage}--\pageref{lastpage}}
\maketitle

\begin{abstract}
Using the DIANOGA hydrodynamical zoom-in simulation set of galaxy clusters, we analyze the dynamics traced by stars belonging to the Brightest Cluster Galaxies (BCGs) and their surrounding diffuse component, forming the intracluster light (ICL), and compare it to the dynamics traced by dark matter and galaxies identified in the simulations. We compute scaling relations between the BCG and cluster velocity dispersions and their corresponding masses (i.e. $M_\mathrm{BCG}^{\star}$-- $\sigma_\mathrm{BCG}^{\star}$, \m200-- $\sigma_{200}$, $M_\mathrm{BCG}^{\star}$-- \m200, $\sigma_\mathrm{BCG}^{\star}$-- $\sigma_{200}$), we find in general a good agreement with observational results. Our simulations also predict $\sigma_\mathrm{BCG}^{\star}$-- $\sigma_{200}$ relation to not change significantly up to redshift $z=1$, in line with a relatively slow accretion of the BCG stellar mass at late times. We analyze the main features of the velocity dispersion profiles, as traced by stars, dark matter, and galaxies. As a result, we discuss that observed stellar velocity dispersion profiles in the inner cluster regions are in excellent agreement with simulations. We also report that the slopes of the BCG velocity dispersion profile from simulations agree with what is measured in observations, confirming the existence of a robust correlation between the stellar velocity dispersion slope and the cluster velocity dispersion (thus, cluster mass) when the former is computed within $0.1 R_{500}$. Our results demonstrate that simulations can correctly describe the dynamics of BCGs and their surrounding stellar envelope, as determined by the past star-formation and assembly histories of the most massive galaxies of the Universe. 
\end{abstract}




\section{Introduction}

Brightest cluster galaxies (BCGs) are a peculiar family of objects: being the most luminous (and most massive) galaxies in the Universe, they are often located at the bottom of the gravitational potential of galaxy clusters \citep[e.g.][]{cui2016does}. Because of their privileged position, their properties are severely influenced by, and in turn heavily affect, the extreme environmental conditions of cluster centers, the latter being sites of interesting evolutionary phenomena (e.g. dynamical friction, mergers, galactic cannibalism). Furthermore, the evolution of the host halo is thought to be tightly connected to that of the BCG, since their mutual alignment is predicted to be present since at least $z\leq 2$ \citep{ragone2020evolution}. BCGs often host the largest supermassive black holes in the Universe, whose presence is usually responsible for intense star formation histories due to the interplay between active galactic nuclei (AGNs) and stellar feedback \citep[e.g.][]{chen2018likely}.
Likewise, BCGs represent the dominant population at the massive end of the galaxy luminosity function. It is precisely for these reasons that the resulting properties (e.g. high luminosity, extended stellar envelope, quenched star formation) are influenced both by their large masses and the surrounding cluster environment \citep{von2007special}. Both effects combined make them attractive targets to benchmark models of galaxy formation. 
    
To understand how the hierarchical growth of collapsed halos regulates the observational properties of the galaxies hosted in such halos, we often rely on the correlation between the properties of the central galaxy and the host dark matter (DM) halo. Among these, the correlation between the BCG stellar mass $M_\mathrm{BCG}^{\star}$ and the cluster mass has been studied over the years in both observational  \citep[e.g.][]{bellstedt2016evolution,kravtsov2018stellar,erfanianfar2019stellar} and theoretical works \citep[e.g.][]{bahe2017hydrangea,pillepich2018first,ragone2018bcg}. Indeed, several theoretical models based both on semi-analytical models of galaxy formation \citep[e.g.][]{de2007hierarchical} and hydrodynamical simulations \citep[e.g.][]{ragone2018bcg,pillepich2018first} suggest that the mass growth of the BCGs at late times is controlled by the hierarchical accretion rather than in situ star formation. Nevertheless, the debate on the formation mechanism of BCGs is not settled yet, while only recently simulations have started to provide predictions in agreement with the observational measurements \citep{ragone2018bcg}. 

As a further line of investigation, recent studies have concentrated on the observational study of stellar velocity dispersion in BCGs, as a way to obtain information on the dynamical history of their assembly \citep[e.g.][]{remus2017outer,sohn2020velocity,bose2020measuring,sohn2021}. \cite{sohn2020velocity} investigated the relationship between the stellar velocity dispersion of the BCG and the cluster velocity dispersion for a large spectroscopic sample. The observed correlation is tight and velocity dispersion measurements are less affected by systematics which instead are inevitably introduced in estimates of the stellar mass of the BCG. More recently, \cite{sohn2021} confirmed that the ratio between the BCG velocity dispersion \dispBCG and the overall velocity dispersion of clusters \dispCL is a steady decreasing function of \dispCL, in contrast with predictions from cosmological hydrodynamical simulations by \cite{dolag2010dynamical} and \cite{remus2017outer}. In any case, it is clear that constraining this relation will open up more possibilities to investigate BCG properties and correlation with the host cluster. 

Although promising, analyses of the BCG velocity dispersions pose several issues which need to be taken into account. Part of the challenge resides in the different approaches followed to estimate the BCG properties in observations and simulations. These include, but are not limited to, the choice of aperture and the different selection criteria of the galaxy in cluster regions (e.g. most luminous or most massive galaxy). To complicate the comparison, the evolution of the BCG is tightly connected to the build-up of a diffuse stellar envelope, the so-called intracluster light (ICL), whose properties are rather unique. The ICL is predicted to form at relatively late times (primarily from tidal stripping and galaxy mergers) and to be gravitationally bound to the cluster as a whole rather than to a single galaxy \citep[e.g.][]{murante2007importance,contini2014formation,contini2018different,montes2018intracluster,montes2019intracluster}. The spatial extent of this component, which overlaps the BCG, can reach up to hundreds of kpc, similarly to the DM in galaxy clusters \citep{pillepich2018first,montes2019intracluster}. However, determining the amount and extent of the ICL component from observations is quite challenging, as it requires both wide and deep observations, capable of capturing spatially extended low-surface brightness regions \citep[in MACS J1206.2-0847][]{presotto2014intracluster}. The comparison of the observational analyses is further complicated by the use of different apertures to define the BCG stellar mass that can lead to the inclusion of a more or less important fraction of the ICL component. A consistent choice among different studies on including the ICL stars may alleviate the tension observed in the estimated evolution of the BCG stellar mass with redshift \citep{zhang2016galaxies,ragone2018bcg}.
   
Moreover, BCGs also offer a valuable route to study the core cluster of host halos, where the DM distribution is a key prediction in cosmological models (see the seminal papers by \citealt{navarro1996cores,navarro1997universal}; more recently \citealt{he2020constraining}). However, standard methods to determine accurate mass estimates of cluster halos (X-ray combined with Sunyaev-Zeldovich effect, and weak lensing measurements) often fail to probe the innermost regions of clusters. X-ray measurements of the intracluster medium are sensitive to the presence of cool front and temperature fluctuations near the cluster center \citep{arabadjis2004extracting}, additionally they are affected by systematic uncertainties due to deviations from hydrostatic equilibrium, albeit especially in the outer regions \citep{rasia2006systematics}. On the other hand, in weak lensing surveys, clusters are imaged deeply to obtain a resolved lensing signal but usually only cover limited areas around each cluster. This issue is present especially at low redshifts where this regime requires imaging of greatly high power of resolution \citep[e.g.][]{joffre2000weak}.

On the other hand, the stellar kinematics of BCGs provides us with information on the total mass at small radii, given that the stellar velocity dispersion profile directly relates to the total gravitational potential well (and thus, the mass distribution) in which stars are moving. Indeed, \citet{sartoris2020clash} performed a dynamical analysis in the core region of Abell S1063, via the MUSE integral field spectroscopy. These authors were able to reconstruct the inner logarithmic slope of the DM density profile, inferring the latter from solving the spherical Jeans equation with the stellar velocity dispersion and the projected phase-space distribution of the other cluster galaxies, to find a value consistent with the cold DM model predictions. These results were in contrast with previous observational measurements, based on a combination of strong lensing and BCG kinematics \citep[i.e.][]{newman2013density}, thus stimulating a discussion on the possible reasons behind this discrepancy \citep{he2020constraining}. In general, we expect stellar velocity dispersion profiles of BCGs to echo gravitational mechanisms operating at the cluster centers. However, as the size of the spectroscopic samples has been increasing, we have come across a large variety in the slopes of the velocity dispersion in the centers of BCGs \citep{loubser2018diversity,loubser2020dynamical}. Whether this diversity of slopes is due to the differences in the mass profiles of the galaxies \citep{barnes2007density} or their different evolutionary paths, this is an open question whose answer requires a better understanding of its dynamical origin.

The analysis presented in this paper aims to examine the dynamics of the distinct tracers of matter in galaxy clusters – namely DM, stars, and galaxies – with a particular interest in characterizing the stellar population of the BCGs. This is addressed by analyzing the DIANOGA set of cosmological hydrodynamical simulations of galaxy clusters \citep{bassini2020dianoga, marini2021phase}. More specifically, the main objectives of this work can be summarized as follows: (i) testing the reliability of state-of-art simulations in describing the observed dynamical properties of galaxy clusters as traced by BCG stars and ICL in the central regions and by the overall cluster galaxy population; (ii) examining the dynamics of BCGs in relation to the host cluster properties and analyze the principal features of the (stellar and non-stellar) velocity dispersion profiles. This is carried out through the analysis of simulated galaxy clusters at two different resolutions, which has the twofold advantage to achieve the maximum information available and to test for numerical convergence. 

This paper is organized as follows: in Section \ref{sec:simulations} we briefly present the basic characteristics of the simulation set analyzed; Section \ref{sec:observables} illustrates the definitions adopted to identify the observable properties of the BCGs and the galaxy clusters in our simulations; in Section \ref{sec:connection} we discuss our understanding of the dynamics of the BCGs in relation to the evolution of the cluster as a whole;  Section \ref{sec:dispersion} examines simulated velocity dispersion profiles and compares them to observational results. Finally, Section \ref{sec:conclusions} summarizes the main results. 

\section{Simulations}
\label{sec:simulations}

The DIANOGA set of simulated galaxy clusters has already been presented in previous papers \citep[][and references therein]{ragone2018bcg,bassini2020dianoga, ragone2020evolution, marini2021phase} to which we refer for additional details. In this section, we outline the main features of these simulations, focusing on those aspects which will be of interest in this analysis.

Simulations were performed with a custom version of GADGET--3, a Tree/Particle Mesh Smoothed-Particle-Hydrodynamics (SPH) code \citep{springel2005cosmological}. From a given set of initial conditions, we carried out hydrodynamical simulations at two different levels of resolution \citep[see Table 1 in][]{marini2021phase} to test for numerical convergence: we name \ll'Hydro-1x\rr' the base (DM particle mass $8.3\times10^{8}$ h$^{-1}\, M_{\odot}$ and initial gas particle mass $3.3\times10^{8}$ h$^{-1}\, M_{\odot}$) and \ll'Hydro-10x\rr' the high (DM particle mass $8.3\times10^{7}$ h$^{-1}\, M_{\odot}$ and initial gas particle mass $3.3\times10^{7}$ h$^{-1}\, M_{\odot}$) resolutions respectively. The Plummer equivalent gravitational softening adopted for DM particles in the \ll'Hydro-10x\rr' run is 1.4 h$^{-1}$ kpc (3.75  h$^{-1}$ kpc in the 1x) while the softening lengths for gas, star, and black hole particles are 1.4, 0.35, and 0.35 h$^{-1}$ kpc respectively (3.75, 1.0, 1.0 h$^{-1}$ kpc in the 1x). The adopted cosmological parameters are $\Omega_M=0.24$, $\Omega_b =0.04$, $n_s = 0.96$, $\sigma_8=0.8$ and $H_0 = 72$ km s$^{-1}$ Mpc$^{-1}$. 

Additionally, our sets of simulations incorporate a treatment of the unresolved baryonic physics including sub-resolution models for star formation and galactic outflows driven by supernova feedback according to the original model in \cite{springel2003cosmological} while metal enrichment and chemical evolution, whose stellar yields are specified in \cite{biffi2018enrichment,biffi2018origin}, follow the formulation described in \cite{tornatore2007chemical}. The AGN feedback is implemented as described in Appendix A of \cite{ragone2013brightest} with a new prescription for the coupling energy to the gas particles \citep[][]{planelles2013baryon,planelles2013role}.

At the base resolution, the set of simulated clusters were extracted from 29 lagrangian regions in a parent N-body box of size 1 h$^{-1}$ Gpc. Each region was then re-simulated at higher resolution with the inclusion of baryonic physics. Among these halos, 24 have mass $M_{200}>8 \times 10 ^{14}\, {\rm h}^{-1} M_{\odot}$ while 5 isolated ones have masses in the range $1-4 \times 10 ^{14}\,{\rm h}^{-1} M_{\odot}$. The Hydro-10x sample includes eight lagrangian regions from which we select, among the ten most massive subhalos, those that are uncontaminated by low-resolution DM particles within \r200\footnote{We define \r200 as the radius encompassing a mean overdensity equal to $\Delta_{200}$ times the critical cosmic density $\rho_c(z)=3H^2(z)/8\pi G$.}. Our final sample is selected in the Hydro-10x from among each lagrangian region, as the main halo of the FoF groups with total DM mass (defined as the sum of all the DM particles bound to the group) larger than $5\times10^{9}$ h$^{-1}\, M_{\odot}$, this comprises of 57 halos. At the base resolution, we only select the main halo from the largest FoF groups, for a total of 29 halos.

\subsection{Identification of substructures}
\label{subsec:subfind}
The preliminary step to classify substructures in cluster halos is to run, in each high-resolution Lagrangian region, the Friends-of-Friends (FoF) algorithm to arrange particles in groups with a linking length of 0.16 the mean interparticle separation. Substructures are then identified by applying the Subfind algorithm \citep{dolag2009substructures} on each FoF group. We consider a substructure to be resolved if it includes a minimum of 50 (DM or stellar) particles and if it has a minimum DM mass of $5\times 10^9 M_{\odot}$ in the Hydro-10x run ($5\times 10^{10} M_{\odot}$ in the Hydro-1x case). Identified substructures that fulfill these criteria are considered as "bona fide" cluster galaxies. 

In addition to substructures, Subfind also identifies the central halo and all the stars belonging to it. In its original formulation, Subfind does not distinguish between stars belonging to a well-identified BCG and stars belonging to the ICL. 

\subsection{Separating BCG and ICL stellar components}
\label{subsec:stellarsubfind}
In order to overcome this limitation of Subfind, we implement two different algorithms to determine the dynamical distinction between ICL and BCG in the stellar envelope of the main halo of each group. 

\begin{figure}
    \centering
    \includegraphics[scale=0.35,angle=0.0]{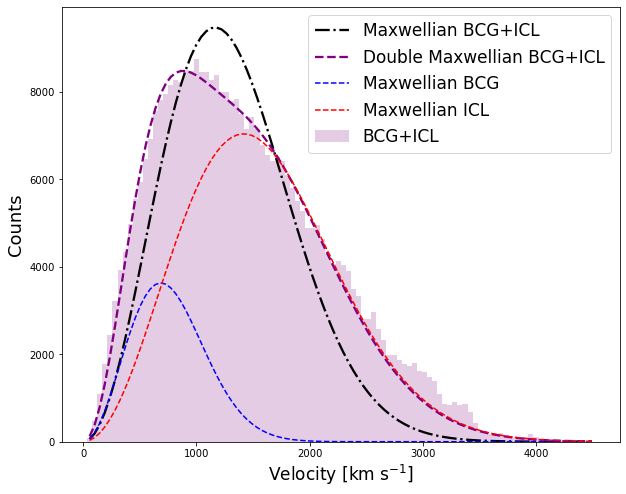} 
    \caption{Velocity histogram (purple) of the main halo stellar component (BCG+ICL) of one of the clusters (\m200 $= 1.65\times 10^{15}$\msun) and a double Maxwellian fit to it (dashed purple line). Red and blue dashed lines are showing the individual Maxwellian distribution which are associated to the unbound ICL ($\sigma_\textrm{ICL}^{\star} = 1417$  km s$^{-1}$) and bound BCG ($\sigma_\textrm{BCG}^{\star} =685$  km s$^{-1}$) components respectively. The black dashed line is the best-fit to the ICL+BCG particles with a single Maxwellian. }
    \label{fig:double_maxwellian}
\end{figure}

The first one is a modified version of Subfind, as presented by \cite{dolag2010dynamical} that relies on the assumption that the two velocity distributions of stars belonging to the ICL and the BCG have each a Maxwellian shape so that the overall velocity distribution of stars is described by a double Maxwellian,
\begin{equation}
    N(r) = k_1 \nu_1^2\exp\left(-\frac{\nu_1^2}{2\sigma_1^2}\right)+k_2 \nu_2^2\exp\left(-\frac{\nu_2^2}{2\sigma_2^2}\right)
\end{equation}
Figure \ref{fig:double_maxwellian} shows the particle velocity distributions of the stars in the main halo of one cluster (\m200 $= 1.65\times 10^{15}$\msun) at $z=0$. We observe that the single Maxwellian (best-fit $\sigma = 1164$ km s$^{-1}$), represented by the dashed black line, does not provide a good fit to the particle distribution. On the other hand, when the fitting procedure is attempted with a double Maxwellian (purple dashed line) the agreement is much more evident, and the residuals can be considered negligible. The diffuse ICL is associated with the Maxwellian with the larger velocity dispersion ($\sigma^{\star}_\textrm{ICL} = 1417$ km s$^{-1}$), in contrast, the BCG, having colder dynamics, populates the distribution at lower dispersion ($\sigma_\textrm{BCG}^{\star} = 685$ km s$^{-1}$). Furthermore, we tested that a triple Maxwellian does not improve the results. To assign each stellar particle to either one of the two dynamical components, the algorithm follows an unbinding procedure, by iteratively computing the gravitational potential given by all particles within a certain radius. This identifies two stellar populations, classified as "bound" and "unbound", which are separately fitted by a single Maxwellian distribution. If the best-fit parameters of the initial double Maxwellian coincide with those obtained in the single Maxwellians, then the procedure is completed, otherwise, the radius of the sphere is changed accordingly and the computation is remade. Hence, by measuring differences in the dynamics we spatially separate the diffuse ICL and the stars bound to the BCG. This first technique is applied to the Hydro-1x set of simulations.
The second algorithm is a novel one and exploits machine learning methods to identify the two dynamical components. This method is described in detail in a forthcoming paper (Marini et al. 2021, in preparation). A brief description is provided in Appendix \ref{AppendixB}, where we illustrate a few key concepts on which it is based. The algorithm is a  supervised Random Forest classifier \citep{breiman2001random} trained on the results provided by Subfind when applied to the clusters in the Hydro-1x set to separate BCG and ICL stellar populations. For each stellar particle in the main halo, the classifier takes as input vector the cluster mass \m200, the information on the radial distance from the center of the cluster and the particles velocity, while it gives as output the corresponding label (ICL or BCG). The main reasoning behind using these specific features can be found in the technique employed by Subfind, since it is mainly based on an unbinding procedure supported by both particles' spatial and velocity distribution. To estimate the accuracy of the algorithm, we define the accuracy score between the predicted $\hat{y}$ and true $y$ label as
\begin{equation}
    \textrm{accuracy}(\hat{y},y) = \frac{1}{N-1}\sum_{i=0}^{N}{\delta(\hat{y_i}-y_i)}
\end{equation}
where N is the number of objects classified and $\delta$ is the Dirac delta function. Employing this definition, the algorithm is estimated to be accurate at $\sim$ 80-90 percent. We apply this technique to the star particles in the main halos of the cluster in the Hydro-10x.

\section{Observables in simulations}
\label{sec:observables}
\begin{figure*}
    \centering
    \includegraphics[scale=0.45,angle=0.0]{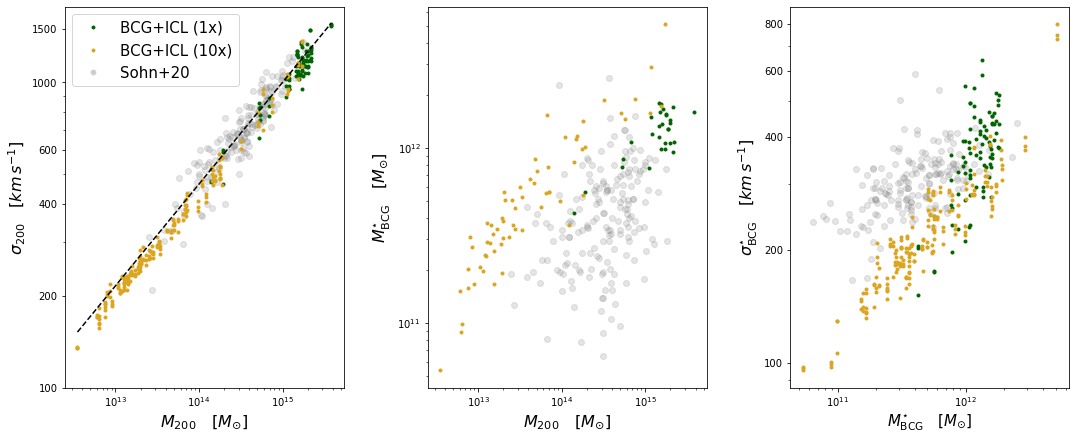}
    \caption{A comparison of some of the general properties of our simulated set of clusters with the observational results by \protect\cite{sohn2020velocity}. \emph{Left panel:} scaling relation between cluster velocity dispersion, $\sigma_{200}$ and total cluster mass, $M_{200}$; \emph{central panel:} scaling relation between BCG stellar mass, $M_\textrm{BCG}^{\star}$ and $M_{200}$; \emph{right panel:} scaling relation between BCG velocity dispersion, $\sigma^{\star}_\mathrm{BCG}$ and $M_\textrm{BCG}^{\star}$. Legend is common to all panels: dark green and golden points indicate respectively the Hydro-1x and Hydro-10x samples using all the stars in the main halo enclosed in a bi-dimensional aperture of 50 kpc. The gray circles stand for the observational results by \protect\cite{sohn2020velocity}. Velocity dispersions are measured for each simulated cluster within three orthogonal projections.}
    \label{fig:scalings}
\end{figure*}

Here below we briefly describe the different ways we used to characterize BCG and cluster properties in our simulations. Given that our analysis mainly focuses on mass and velocity dispersion, we will provide their definitions in this Section. 
    \subsection{Dynamics of the BCGs}
    Distinguishing the BCG from the stellar diffuse envelope within clusters is notoriously non-trivial. Although BCGs are usually the brightest (and most massive) galaxies in their cluster, positioned in a privileged position at the bottom of the gravitational potential, many factors contribute to complicating the analysis, especially in observations. In some cases, the central galaxy may not coincide with the brightest galaxy in the cluster, or its position may be shifted with respect to the geometrical center of the cluster halo \citep[e.g.][]{de2021brightest}. This adds up to the difficulties of defining physical boundaries to the galaxy, given that its stellar population is wrapped in a diffuse stellar component, the so-called ICL, which is gravitationally bound to the cluster potential. A complete discussion on the methods to discern the two is beyond the scope of this paper, we refer the interested reader to \cite{rudick2011quantity} for a comprehensive discussion.
        
    To account for possible shifts between the BCG center of mass and the center of the halo (identified by Subfind as the position of the particle with the minimum in the gravitational potential), we iteratively compute the stellar center of mass within $0.1 R_{500}$, centering it at each iteration on these coordinates and recomputing it within $0.1 R_{500}$ from this new center. Avoiding this correction causes the measurement of (non-physically) large stellar velocity dispersions in several clusters. Furthermore, given that our simulations provide us with the full phase-space information on the distribution of particles, we employ for each simulated cluster the projections along the three orthogonal directions.

    In the analysis of observed clusters, all the properties of the BCG are measured within a circular aperture centered on the brightest galaxy member in a given band \citep[e.g][]{sohn2020velocity} or on the galaxy closest to the X-ray peak \citep{loubser2018diversity}. The choice of the aperture radius is somewhat arbitrary in the literature, still masses measured within a fixed physical radius allow us to neatly compare predictions from simulations with observations. We choose to identify all BCG observables as the properties yielded by the star particles in a cylinder long \r200 centered on the BCG and within a 50 kpc physical radius (we note that by reducing the aperture to 30 kpc we do not expect major differences in the BCG properties out to $z \simeq 1.5$; \cite{ragone2018bcg}). We select only stars in the main halo, excluding those which Subfind identifies as bound to other substructures, and we measure the physical properties within this circular area. We point out that with this method we include not only the dynamical information on the BCG but also the contribution from the ICL. In the following, we will refer to this definition as \ll'BCG+ICL\rr'.

    As anticipated in Section \ref{subsec:subfind}, we include a second definition of the BCG, based on the dynamical distinction between BCG and ICL in the stellar population of the main halo hosted in galaxy clusters. Although this technique is applicable only in simulations, it provides us with a substantial understanding of the assumptions made when we take the BCG in a fixed aperture instead of retrieving it in a more physically motivated way. For consistency, we compute all physical properties within a cylinder long \r200, centered on the BCG, and within a 50 kpc aperture, this time including only stars which are dynamically associated with the BCG. We will refer to this second definition as \ll'BCG\rr' or \ll'BCG-only\rr' interchangeably. 
 
    Therefore, we define the line-of-sight (LOS) stellar velocity dispersion $\sigma_\mathrm{BCG}^{\star}$ as the r.m.s. of the velocity distribution of the star particles (identified as BCG+ICL or BCG-only) within the cylindrical projection. We find the r.m.s. estimator to obtain stellar velocity dispersions to closely agree with the observational data. All the observational datasets employed in this analysis comprise velocity dispersion measurements recovered from the galaxy spectra with the Gauss‐Hermite series\citep{cappellari2004parametric}.
    
    We note that using the biweight estimator (which, on the other hand, is used to compute the total cluster velocity dispersion) does not guarantee the same level of agreement. Additionally, velocity dispersions are measured within the three orthogonal projections. The BCG stellar mass $M_\mathrm{BCG}^{\star}$ is computed as the median of the sum of the selected star particles within the same bi-dimensional aperture. 

    \subsection{Dynamics of the clusters}
    
    Observed cluster properties are usually retrieved from the dynamics of galaxy members. In our simulated clusters, {\em bona fide} galaxies correspond to gravitationally bound substructures, which we identify through the SubFind algorithm (see Section \ref{subsec:subfind}). However, given that substructures in simulations sometimes fail in reproducing the phase-space structure of the real galaxy population in clusters \citep[e.g.][]{hirschmann2016galaxy,marini2021phase}, it is advisable to also employ DM particles to characterize the cluster dynamical properties. These particles have been selected from the main halo of the central group, excluding the contribution from those bound to substructures. In this regard, the dynamics followed by DM particles should be a robust tracer of the cluster potential, unlike substructures that are dependent on the Halo Finder, and possibly also suffering from the effect of dynamical friction. In Section \ref{sec:connection} we further comment on the impact of choosing DM particles with respect to galaxies.
    
    Following the observational approach, we select DM particles within a cylinder long \r200 whose base of radius \r200 is centered on the cluster center, corresponding to the position of the particle with the minimum value of the gravitational potential. We determine the DM particles kinematics (i.e. the cluster velocity dispersion $\sigma_{200}$) with the biweight estimator. Given that galaxies are most likely affected by fore-/back-ground interlopers in the observational analysis, observers often choose to use the biweight because it underweights the tails of the velocity distribution which are expected to be populated by outliers \citep[][]{beers1990measures}. Analogously, we define the cluster mass $M_\mathrm{200}$ as the total particle mass within the same cylinder.
    
\section{Scaling relations between BCGs and clusters}
\label{sec:connection}
Among the observational properties of BCGs, that have been studied over the years, the scaling relation between the BCG stellar mass $M_\mathrm{BCG}^{\star}$ and the host halo mass $M_{200}$ holds a special place given its direct connection with the hierarchical growth of structure. We expect both masses to depend on a combined action of different physical processes which include (but are not limited to) halo assembly history, gas accretion, AGN feedback, and galaxy mergers. 

In the following, we will present several properties of our set of simulated clusters, and compare them with recent observational results by \citet{sohn2020velocity}. These authors analyzed the HeCS-omnibus cluster sample which includes 227 objects (\m200 $=2.5-18.4\times 10^{14}$ \msun) observed from a combination of both photometric and spectroscopic surveys, over the redshift range $0.02\leq z\leq 0.3$. Cluster masses \m200 are measured via caustic method \citep{diaferio1997infall,serra2013identification}, while stellar masses are estimated with synthetic spectral distribution (SED) model fitting  ($M_\textrm{BCG}^{\star}$ ranges $6.46-251.19 \times 10^{10}$ \msun). The authors investigated several properties connecting the BCG with the hosting DM halo, in particular, by measuring the correlation between the central stellar velocity dispersion of the BCG with the cluster velocity dispersion, finding it to be remarkably tight. One of the main questions left open by \citet{sohn2020velocity} is whether state-of-art numerical simulations are capable of reproducing similar results, a question which we will address in this section. In the first part, we will focus on examining the differences between the observational dataset and the results from our simulations in the approach closest to the observational procedure (namely, the BCG+ICL sample). In the second part, we will investigate the effects of excluding the ICL from the stellar component. 

    \subsection{Scaling relations}
    \begin{figure}
        \centering
        \includegraphics[scale=0.45,angle=0.0]{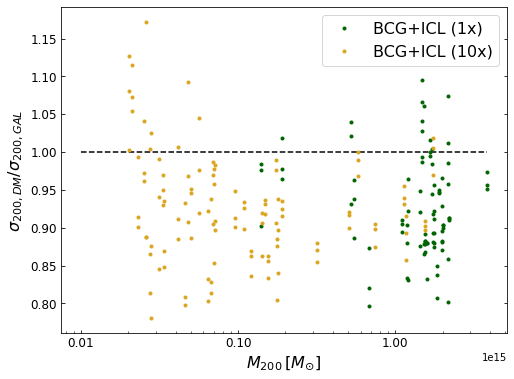}
        \caption{The velocity bias as a function of cluster mass in the Hydro-1x (dark green points) and Hydro-10x samples (golden points). The dashed black line reports the line for no bias (\dispCL/\dispBCG$=1$). Velocity dispersions are measured for each simulated cluster within three orthogonal projections.}
        \label{fig:bias_vs_m200}
    \end{figure}
    \begin{figure}
        \centering
        \includegraphics[scale=0.45,angle=0.0]{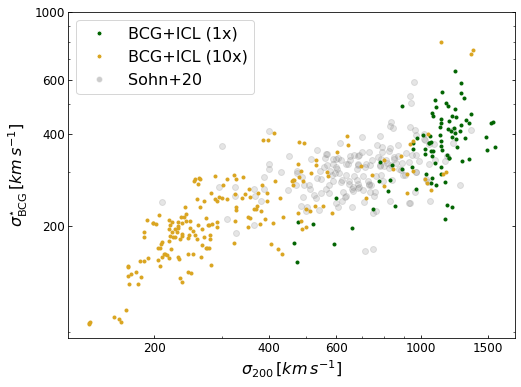}
        \caption{Relationship between BCG and cluster LOS velocity dispersion from clusters in  \protect\cite{sohn2020velocity} (in grey), and at $z=0$ the Hydro-1x and Hydro-10x simulations. The colored points are the result of selecting all stars in the main halo within the cylindrical projection. Velocity dispersions are measured for each simulated cluster within three orthogonal projections.}
        \label{fig:Sohn+2019-gr0Gauss}
    \end{figure}
    
    Figure \ref{fig:scalings} illustrates (from left to right panel) \dispCL vs \m200, $M_\mathrm{BCG}^{\star}$ vs \m200 and $\sigma_\mathrm{BCG}^{\star}$ vs $M_\mathrm{BCG}^{\star}$ from the cluster sample in our simulations at $z=0$ compared to the observed population provided by \cite{sohn2020velocity}. In this regard, we decided to employ the simulations at $z=0$, given that in any case, our results do not significantly change at late times. Legend is common to all panels: dark green and golden points indicate respectively the Hydro-1x and Hydro-10x samples using all the stars in the main halo enclosed in a bi-dimensional aperture of 50 kpc (BCG+ICL sample). We remind that the BCG+ICL sample is the sample selected with an approach closest to the observational one and shall be regarded as the right sample for comparison. The gray circles stand for the BCG measurements studied by \protect\cite{sohn2020velocity}. More in detail, in the left panel, we examine the relationship between the cluster LOS velocity dispersion \dispCL and cluster mass $M_{200}$, to understand the capability of our simulations to recover this well-constrained correlation \citep[e.g.][]{evrard2008virial,saro2013toward}. Indeed, once calibrated, the cluster velocity dispersion is an observable extremely sensitive to the cluster mass. We find that DM particles trace a relation fully consistent with both the theoretical expectation, a power-law with slope 1/3 (black dashed line), and the observational data. Given this tight correlation, in the following, we will often refer to \ll'high-mass\rr' or \ll'low-mass\rr' cluster samples as those with high or low-velocity dispersion values respectively. We point out that the rationale of choosing the DM particles as tracers of the internal dynamics of simulated clusters, instead of using the galaxies identified as substructures, lies in the fact that substructures suffer from a velocity bias possibly larger than in real clusters. In this regard, Figure \ref{fig:bias_vs_m200} shows the dependence of the velocity bias with halo mass \m200 in the same sample of clusters. In our simulations, we measure a consistent velocity bias of the substructures with respect to the DM particles of $\sim 10$ percent at all selected masses which is in line with previous measurements \citep[e.g.][]{lau2009effects,munari2013relation} 
  
    As for the central and right panels, we verify whether the BCG properties in simulations have realistic values compared to the observational data. In the central panel, we examine the relation between $M_\mathrm{BCG}^{\star}$ and cluster mass $M_{200}$, a relation generally known to be difficult to reproduce in simulations, as BCG stellar masses are sensitive to the details of the feedback processes. We note that \citet{bassini2020dianoga} and \citet{ragone2018bcg} previously discussed at length these same results on both resolutions, finding that the Hydro-10x set tends to overestimate the stellar mass of the BCG with respect to the observational data. On the contrary, the Hydro-1x clusters have BCG stellar masses consistent with the observed ones. As discussed by \citet{bassini2020dianoga}, rather than a resolution effect, the difference in the BCG stellar masses between the Hydro-10x and Hydro-1x Dianoga clusters lies in the different implementation of AGN feedback. Furthermore, in \citet{ragone2018bcg} the \m200 vs $M_\mathrm{BCG}^{\star}$ relation is compared also to other simulation results, presented by other groups \citep[][]{bahe2017hydrangea,hahn2017rhapsody,pillepich2018first}, yielding comparable results. We note that this discrepancy is larger at the lower masses. 
    
    On the other hand, the BCG stellar mass in the Hydro-1x is compatible with the measurements by \cite{sohn2020velocity}. The right panel shows a tension between the $M_\textrm{BCG}^{\star}$-- $\sigma_\textrm{BCG}^{\star}$ scaling relation for observed and simulated clusters, at least for the Hydro-10x set. Among other factors, an overestimate of the BCG stellar mass contributes to a shift towards the right for the scaling relation predicted by our simulations. In the same plot, we note the presence of three isolated points derived from a cluster in a Hydro-10x simulation with \dispBCG$>$ 600 km s$^{-1}$. These points correspond to the three orthogonal projections of the LOS velocity dispersion of the same cluster. To understand the origin of this outlier, we analyze its recent evolution. These extreme features can be traced back to a sequence of major mergers at late times ($z\leq 0.5$). 
    By studying the velocity distribution of the star particles associated with the main halo we note that this structure did not yet reach dynamical equilibrium by $z=0$, with the distribution showing two separate peaks: a central one, corresponding to the most massive subhalo, and a smaller one, corresponding to the merging structure. The reason why this is particularly evident in the Hydro-10x case, and not in the Hydro-1x, is twofold. Firstly, it is due to how Subfind assigns particles to one halo or the other at two resolutions. Secondly, as the resolution changes the details of the timing of the merging also slightly change. We point out that this feature had already emerged in \cite{marini2021phase}, where we computed the scaling relation with the integrated pseudo-entropy and compare it to the one with the velocity dispersion.

    \subsection{The \dispBCG -- \dispCL ~relation at $z=0$}
    We note that the stellar masses of the BCGs and the total cluster masses are not directly observable quantities, instead, they are inferred from other observable quantities. For instance, scaling relations applied on different observables, such as luminosity \citep[e.g.][]{zhang2011hiflugcs}, temperature in the X-ray \citep{vikhlinin2006chandra} and velocity dispersion \citep{sohn2020velocity} allow to obtain a measure of cluster masses. As for stellar masses, they are often derived by comparing synthetic spectral energy distribution models for single galaxies or through spectroscopy. For this reason, it is useful to also provide a comparison between the stellar velocity dispersion of the BCG and the global cluster velocity dispersion, which are both quantities directly measured from spectroscopic observations. 
    
    In Figure \ref{fig:Sohn+2019-gr0Gauss}, we show the LOS velocity dispersion of the BCG as a function of the cluster LOS velocity dispersion. Our simulations span the full range of velocity dispersion probed by observations, extending to even smaller \dispCL ($<$ 200 km s$^{-1}$) or, equivalently smaller masses. We expect these low--\dispCL galaxies to be isolated early-type galaxies. On the other hand, in the mass range probed by observations (\dispCL $>$ 200 km s$^{-1}$), at least for the case of the Hydro-10x, simulated structures show a rather good agreement with the observed stellar dynamics with a steepening of the relation on the very high-mass end. 
    
    Quite interestingly, the agreement that we find in Figure \ref{fig:Sohn+2019-gr0Gauss} between simulations and observations is not in line with the conclusions reached by \cite{sohn2020velocity} and \cite{sohn2021}, who pointed out a significant tension with the results from the hydrodynamical simulations by \citet{dolag2010dynamical}. In the latter (and more recently, including the galaxy group mass scale, in \citealt{remus2017outer}), the stellar BCG velocity dispersions are derived from the Maxwellian fits of the velocity distribution of BCG and ICL separately. The authors find a self-similar scaling between both BCG and ICL velocity dispersions with cluster mass which suggests the existence of self-similarity between the two components over the entire mass range probed. However, such a self-similar scaling is not retrieved in the observed cluster set. In the following, we will extensively discuss the main differences between our analysis of simulated galaxy clusters and that presented by \citet{dolag2010dynamical} to understand where these differences specifically lie, in the effort to clarify the origin of the previously reported disagreement between observations and simulations.

    \begin{figure}
        \centering
        \includegraphics[scale=0.40,angle=0.0]{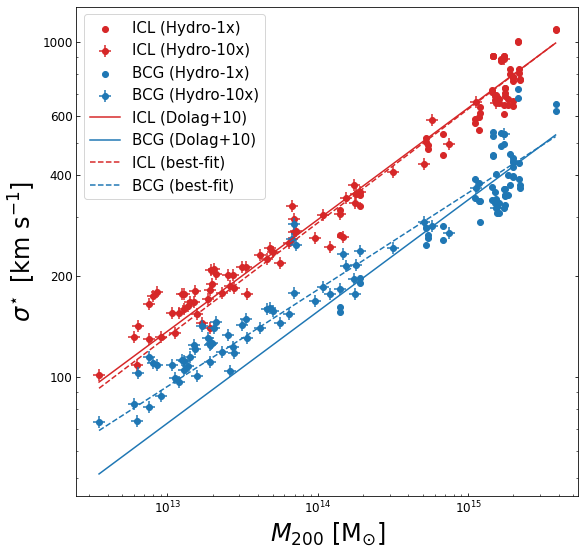}
        \caption{Stellar velocity dispersion of the BCG (blue) and ICL (red) components in the main halo of our set of simulated galaxy clusters, obtained from the double-Maxwellian fit of the particle velocity distribution as a function of host cluster mass \m200. The distinct point styles code the two distinct simulations: simple circles for the Hydro-1x and crossed circles for the Hydro-10x. The solid lines represent the best-fit results from \protect\cite{dolag2010dynamical}, while the dashed lines mark the best-fit from our sample.}
        \label{fig:dispBCG+ICL_vs_m200}
    \end{figure}
	\begin{figure}
        \centering
        \includegraphics[scale=0.4,angle=0.0]{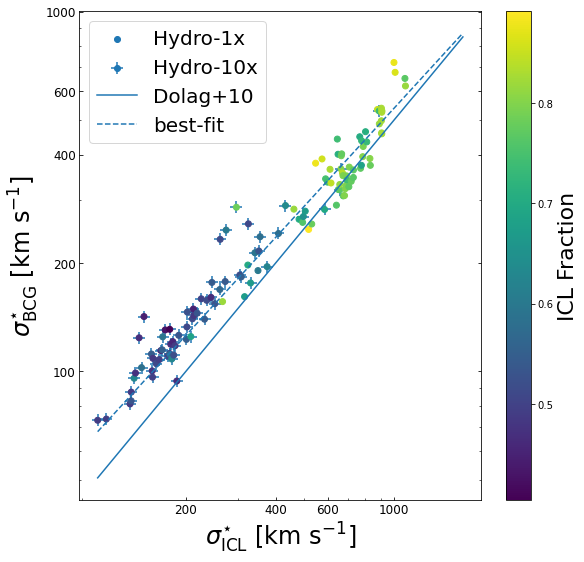}
        \caption{Stellar BCG velocity dispersion vs ICL velocity dispersion in the Hydro-1x (simple circles) and Hydro-10x (crossed circles) runs. Solid line reports the best-fit curve from \protect\cite{dolag2010dynamical} while the dashed line is derived from our cluster set. We use a color-coded description to mark the ICL mass fraction in each cluster.}
        \label{fig:disp_BCG-disp_ICL}
    \end{figure}

    Since the comparison in \citet{sohn2020velocity} is performed between observed velocity dispersions (thus, including ICL) and three-dimensional BCG-only velocity dispersions, as a first step we test whether accounting for projection effects and the inclusion of the ICL contribution affect the results, but these effects turn out to be negligible in our analysis. Therefore, we test whether our cluster sets differ in the virial scaling relations. For this purpose we show the stellar velocity dispersion of BCG (in blue) and ICL (red) as a function of the cluster mass \m200 in Figure \ref{fig:dispBCG+ICL_vs_m200}. The scaling between velocity dispersion and cluster mass is modeled as
    \begin{equation}
        M = A \left(\frac{\sqrt{3}\,\sigma}{10^3 \textrm{km s}^-1}\right)^{\alpha} \times 10^{14} h^{-1} M_{\odot}
    \end{equation} where $\alpha=3$ is value expected from viral equilibrium and it is kept fixed in the analysis by \cite{dolag2010dynamical}. In our analysis, we decided to treat $\alpha$ as a fitting parameter, given that a simple comparison by eye between the virial expectation and our dataset, unlike that from \citet{dolag2010dynamical}, showed a consistent discrepancy in the slope. We distinguish between results from the Hydro-1x (simple circles) and Hydro-10x (crossed circles) simulations to clarify the dependence on resolution. Additionally, we report the best-fit relations as provided by \cite{dolag2010dynamical} (dashed lines) and as obtained from the analysis of our set of simulated clusters (solid lines). The best-fit curves from the two distinct fits in the scaling $\sigma^{\star}_\mathrm{ICL}$-- \m200 are comparable to each other, and our fitting value is very close to the virial expectation, with $1/\alpha\simeq 0.339$. The difference in the normalization is minimal and still within the internal scatter. As for the \dispBCG-- \m200 relation, our BCG set has a significantly different behavior with respect to the best-fit line from  \cite{dolag2010dynamical}, especially when expanding the dynamical range down to the mass range of galaxy groups. It is precisely this difference that explains the difference with respect to \cite{dolag2010dynamical} and \cite{remus2017outer}. A higher BCG velocity dispersion at fixed cluster mass (or global velocity dispersion) in the mass range of galaxy groups brings our measurements in agreement with observational results. This difference is most relevant in the low-mass end, $\sim 5\times10^{13}$ \msun. The slope of our best-fit relation is shallower than the virial value ($1/\alpha\simeq 0.289$) and indicates substantial differences in the mechanisms operating in the BCG accretion with respect to those in the ICL, which traces more the overall virial dynamics of the host halo.
    
    These two effects are consistent with the different dynamical evolution of two stellar components. The ICL is associated with the stellar envelope bound to the cluster potential, and thus we expect this component to dynamically co-evolve with the DM halo, following the self-similar scaling described by the virial theorem. On the other hand, the BCG is expected to form from the dissipative collapse of the gas cooling and fueling star formation within the proto-BCG building blocks, which later assemble into the BCG dry mergers.
    
    Furthermore, we analyzed the correlation \dispBCG-- $\sigma^{\star}_\mathrm{ICL}$. In their analysis, \cite{dolag2010dynamical} found these two velocity dispersions to be proportional to each other, according to
    \begin{equation}
    \label{eq:sigma_BCG-sigma_ICL}
        \sigma^{\star}_\mathrm{BCG} = 0.5\, \sigma^{\star}_\mathrm{ICL}.
    \end{equation}
    We expect this relation not to hold for our set of simulated clusters and groups, given that our \dispBCG distribution is different in the low-mass range. Figure \ref{fig:disp_BCG-disp_ICL} presents our datapoints (circles mark the clusters in the Hydro-1x run, while circled crosses are for the Hydro-10x) with the corresponding best-fit curve (dashed line) for a power-law with varying normalization and slope and the best-fit in \cite{dolag2010dynamical} (solid line). In the light of previous results, it is not surprising that the best-fit slope for our set of simulated clusters is different than 1, as implied by Eq. \ref{eq:sigma_BCG-sigma_ICL}, and the power-law is in fact shallower. Interestingly, we color-coded the data points with the ICL fraction over the total stellar mass associated with the central galaxy, given that \cite{remus2017outer} found that outliers in their cluster distribution were due to clusters with below 5 percent BCG mass fractions. We find a color gradient along the best-fit curve: data points laying on this line have larger ICL mass fractions as we move towards higher $\sigma_\mathrm{ICL}^{\star}$.
    
	In conclusion, we highlighted the origin of the different conclusions reached by \citet{sohn2020velocity,sohn2021} and by our analysis on the comparison between observed and simulated results for the BCG velocity dispersion. Since we expect the dynamics of the BCG to be different from what was observed in \cite{dolag2010dynamical}, the measurements of the stellar velocity dispersions within the innermost regions will also be different, as we have seen in Figure \ref{fig:Sohn+2019-gr0Gauss}. This still holds even if the stellar mass of the BCG in our simulations is overestimated when compared to observational results. Hence, while it is clear that the modeling of BCG stellar mass accretion is extremely sensitive to several different factors, yet the "cold" stellar dynamics of the stars in the central regions is well-reproduced at $z=0$ in our "Hydro-10x" simulations.

    \subsection{Effect of excluding the ICL}
    \begin{figure*}
        \centering
        \includegraphics[scale=0.45,angle=0.0]{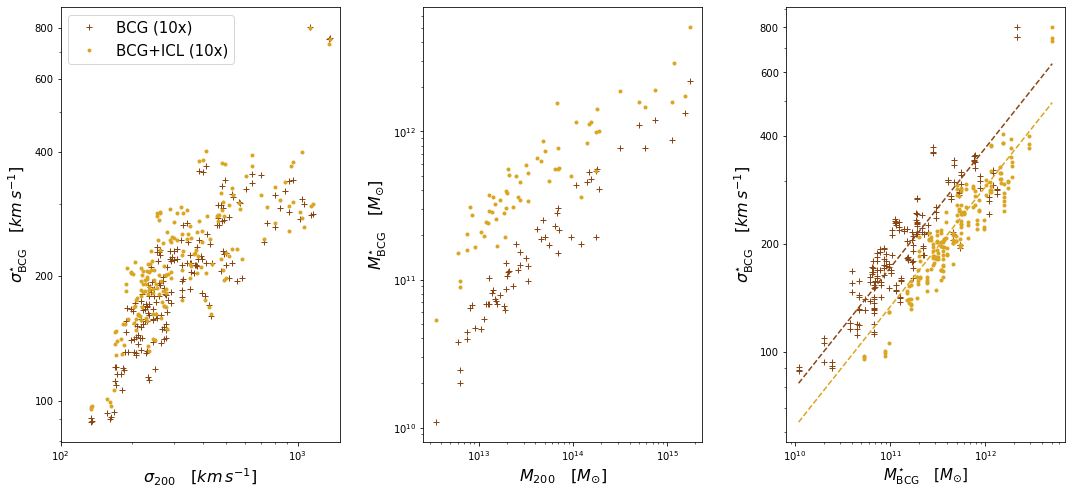}
        \caption{A comparison of the scaling relations for the stellar population of our simulated clusters in both BCG and BCG+ICL definitions, for the Hydro-10x sample. {Left panel:} scaling relation between BCG stellar velocity dispersion, \dispBCG  and total cluster mass, $M_{200}$; \emph{central panel:} scaling relation between BCG stellar mass, $M_\textrm{BCG}^{\star}$ and $M_{200}$; \emph{right panel:} scaling relation between BCG velocity dispersion, $\sigma_\textrm{BCG}^{\star}$ and $M_\textrm{BCG}^{\star}$. In the last panel we also report the predictions from the virial theorem for both samples with colored dashed lines. Legend is common to all panels: golden points indicate the BCG+ICL sample while the brown crosses stand for the BCG sample. Velocity dispersions are measured for each simulated cluster within three orthogonal projections.}
        \label{fig:scalings-checksubfind}
    \end{figure*}
    As a final test, we investigate the effects of excluding the contribution of the ICL to the stellar velocity dispersion of the BCG, according to what is discussed in Section \ref{subsec:stellarsubfind}. This test is based on the BCG sample which, contrary to the BCG+ICL sample, includes only the stars bound to the BCG in a cylinder long \r200, centered on the BCG, and within a 50 kpc aperture. We stress that this test has the only purpose of displaying the impact of removing the ICL with a dynamically motivated procedure, rather than for a comparison with observational results, which cannot separate the effect of the diffuse stellar component to the velocity dispersion measured in projection. In Figure \ref{fig:scalings-checksubfind}, we show the comparison of this dataset with the BCG+ICL sample from the Hydro-10x simulation only, since we checked that results do not change with the lower resolution settings. Legend is common to all panels: brown crosses mark the BCG sample, while golden points label the BCG+ICL sample. From left to right we plot the \dispBCG-- \dispCL, the $M_\textrm{BCG}^{\star}$-- \m200 and \dispBCG-- $\sigma_{200}$ relations respectively. 
    
    The plot in the left panel is complementary to Figure \ref{fig:Sohn+2019-gr0Gauss} and it shows that the velocity dispersions of the "BCG" and "BCG+ICL" cases are different albeit not significantly (within 50 kpc from the center). Similarly, the central panel examines the amount of ICL mass that is still accounted for in the 50 kpc bi-dimensional aperture centered on the central galaxy. Unsurprisingly stellar BCG masses in the BCG-only sample are lower than the BCG+ICL set, given that the former does not take into account the mass contribution from the stellar component assumed to be part of the ICL envelope. This mass difference accounts for a factor $\sim$3. As for the right panel, the stellar velocity dispersion as a function of the stellar mass, we note that the reduction of the stellar mass is only partially compensated by the reduced velocity dispersion, once the ICL contribution to such two quantities is removed. As a consequence, the scaling relation between velocity dispersion and stellar mass has a higher normalization for the "BCG" case. It is striking how the predictions from the virial theorem are still conserved in both cases (a power-law with slope 1/3 as reported by the dashed colored lines).
    
    In conclusion, we find that excluding the ICL component from the 50 kpc aperture on the central galaxy implies lower velocity dispersions, compatible with a cut in dynamically hot stellar population which makes up the ICL, and lower stellar masses.

    \subsection{Redshift evolution of the \dispBCG -- \dispCL relation} 
    
    \label{subsec:redshift}
\begin{figure}
        \centering
        \includegraphics[scale=0.5,angle=0.0]{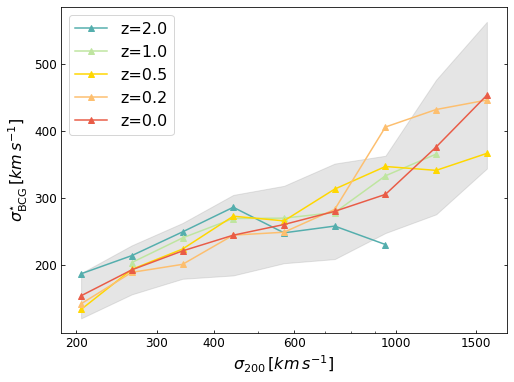}
        \caption{Relationship between LOS BCG velocity dispersion as a function of cluster velocity dispersion, at five different redshifts. Clusters at each redshift are binned in \dispCL. For each bin, we plot the corresponding median value of \dispBCG. The grey shaded band marks the standard deviation of the clusters at $z=0$, with comparable dispersions at different redshifts. At the different redshift, $z=$ 2, 1, 0.5, 0.2, 0, we have the following corresponding number of clusters: 113, 80, 84, 80, and 93. The value of \dispBCG is obtained by also including the contribution of ICL. }
        \label{fig:Sohn+2019redshift_HIST}
    \end{figure}

    Having established that the observed \dispBCG-- $\sigma_{200}$ relation is correctly predicted by our simulations, we investigate its redshift evolution, similar to what is done for the evolution of the stellar mass of BCGs. We select five different redshifts ($z=2$, 1, 0.5, 0.2, 0.0) from the Hydro-1x and Hydro-10x simulations for a total of 113, 80, 84, 80, and 93 clusters at each redshift. For all clusters, the BCG stellar velocity dispersion is computed in projection by including the ICL contribution (BCG+ICL case) with a 50 kpc aperture. At each redshift, we group clusters in bins of \dispCL, with a minimum of 10 clusters per bin. The value of \dispBCG assigned to each bin corresponds to the median value among the clusters belonging to that bin. We report the results of this analysis in Figure \ref{fig:Sohn+2019redshift_HIST}. Additionally, we plot the standard deviation at $z=0$ as a function of the cluster velocity dispersion with the grey shaded band, noting that the standard deviation is similar at all considered redshifts.
    
    We observe that the \dispBCG-- $\sigma_{200}$ relation is almost redshift-independent with no significant variation within its intrinsic scatter. At $z=2$ there is a mild indication for a slightly higher normalization in low-\dispCL halos. This result is possibly consistent with the picture of a rapid BCG formation through dry mergers (i.e. gas-poor and negligible star formation), whose assembly anticipates that of the host halo. 

\begin{figure*}
        \centering
        \includegraphics[scale=0.5,angle=0.0]{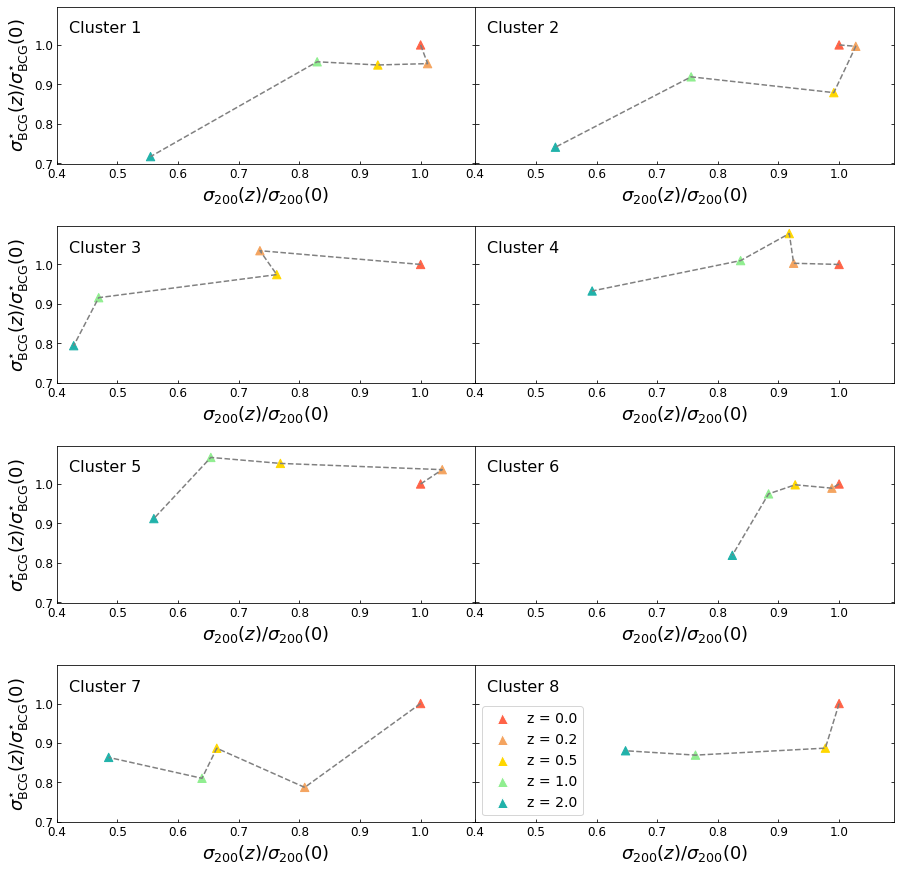}
        \caption{Evolution of the scatter relation between the velocity dispersion of the BCG (normalized at $z=0$) vs the cluster velocity dispersion (normalized at $z=0$) from 8 clusters in the Hydro-10x run. The redshift bins (2, 1, 0.5, 0.2, 0) are labelled with different colors.}
        \label{fig:dispBCG_z0-dispCL_z0}
    \end{figure*}

    As a second test of evolution, we select the most massive FoF group for the eight regions from the Hydro-10x simulation set, and for each of them, we follow the velocity dispersion as a function of redshift up to $z=2$. Figure \ref{fig:dispBCG_z0-dispCL_z0} shows the evolution of \dispBCG vs $\sigma_{200}$ both normalized at $z=0$, with the color-coding indicating the redshift. 
    On average, the cluster velocity dispersion doubles from $z=2$ to $z=0$. The BCG velocity dispersion also increases with time, with most of the evolution occurring at $z>1$. The evolution of the cluster velocity dispersion is rather smooth, although episodes of merger may temporarily cause a sudden increase of \dispCL, followed by relaxation. For instance, this is the case for cluster 5 at $z=0.2$, whose $\sigma_{200}$ is higher than at $z=0$. The same can also happen for \dispBCG, especially if these merging events affect the inner regions. Much like the BCG stellar mass accretion, the stellar velocity dispersion has very little change at late times thus confirming that, once formed through dry mergers, it remains almost independent of the cluster mass growth. In other words, we expect the late-time mass accretion in the central regions not to significantly affect the dynamics of the BCG.

\section{Velocity dispersion profiles}
\label{sec:dispersion} 

Stellar velocity dispersion profiles of the BCGs provide an interesting diagnostic to investigate the inner core of galaxy clusters. On one hand, they allow us to explore the dynamics of the innermost cluster regions, which is expected to reflect the assembly history of the BCG. On the other hand, the combination with strong lensing measurements allows reconstructing the total mass distribution (while also yielding the DM distribution after the subtraction of the visible matter). In this Section, we present the results of our analysis of the BCGs dynamics as traced by the velocity dispersion profiles. This will be performed with a two-steps approach, according to which we will derive the expected velocity dispersion profiles for the distinct tracers in galaxy clusters, while later we will assess the main features that characterize the observed BCG stellar population and quantify the ICL intake in these profiles.

    \subsection{Dynamical properties of cluster core regions}

	    Recently \cite{sartoris2020clash} determined the full dynamical analysis of Abell S1063 ($R_{200}=$ 2.36 Mpc, as inferred from weak lensing measurements) combining two different tracers: the kinematics of the cluster galaxy members and the stellar velocity dispersion profile of the BCG. For this cluster at $z=0.3$, the observations consisted of an intensive spectroscopic campaign using the VIMOS and MUSE spectrographs at the VLT. With our analysis, we want to verify whether our simulations of clusters of comparable size reproduce the cluster dynamics as traced by the galaxy population and, in the innermost region, by the BCG. For this comparison, we extend the aperture radius to include all the stars in the main halo out to \r200. This has the advantage to provide us information on the dynamics traced by both the stellar component of the main cluster halo and by the cluster galaxies at large radii. 
	    
\begin{figure}
            \centering
                    \includegraphics[scale=0.5,angle=0.0]{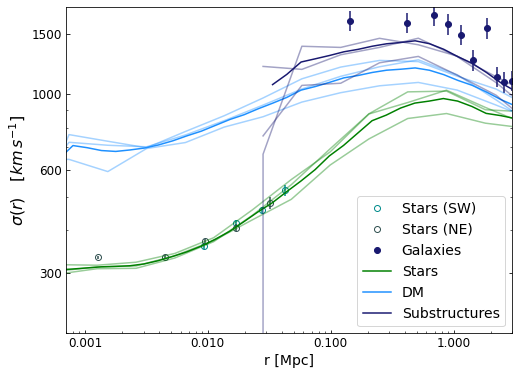}
                    \caption{Comparison with the LOS velocity dispersion from \protect\cite{sartoris2020clash} (points with errorbars) and the projected velocity dispersion profiles from one of our simulated clusters at $z=0$ in the Hydro-10x set. Error bars in the observational points refer to a 68 percent confidence level. The solid lines are the mean of the three projected velocity dispersion profiles (faded lines).}
                    \label{fig:Sartoris+2020}
        \end{figure}

        Figure \ref{fig:Sartoris+2020} shows the comparison between observational results from \cite{sartoris2020clash} and one of the simulated clusters from the Hydro-10x runs at redshift $z=0$. The latter was selected for its closeness in size to the observed cluster (\r200$=$ 2.34 Mpc). Solid lines are for the sample obtained from simulations, while the filled and open circles correspond to the stars and galaxies of Abell S1063, respectively. Error bars on observational data points refer to the reported 68 percent uncertainties. The accuracy of our simulated cluster in reproducing the observed data points is quite remarkable, especially for the velocity dispersion profile of stars. In the outer regions, we see that DM and galaxies profiles from the simulations both capture the observed negative gradients, albeit with a slightly lower normalization for $r < 1$ Mpc. This difference can be due to several factors. Indeed, we stress that the cluster selected from simulations has been chosen for its size (the most massive in our cluster sample and the closest to the observed cluster mass), thus differences in the dynamics of its components can be ascribed to distinct dynamical states or specific cluster formation histories.
        
        \begin{figure}
            \centering
            \includegraphics[scale=0.5,angle=0.0]{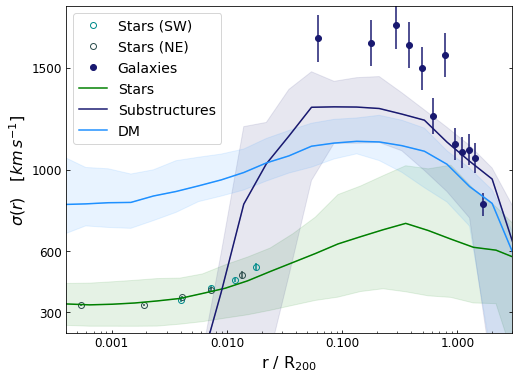}
            \caption{Comparison of the observed LOS velocity dispersion from \protect\cite{sartoris2020clash}, with results from the combined sample of simulated clusters at $z=0$ from the Hydro-10x set. The shaded areas represent respectively the 16\th and 84\th percentile given by the simulated cluster sample.}
        \label{fig:Sartoris+2020-stack}
        \end{figure}

        A further step can be taken by extending the comparison to include the eight clusters from the Hydro-10x set of simulations to infer the general characteristics of the distinct tracers when combining all the simulated clusters. The results are shown in Figure \ref{fig:Sartoris+2020-stack}, with shaded areas showing the 16\th and 84\th percentile given by the entire cluster sample.
        
        Except for particularly disturbed systems, as it has been discussed in Section \ref{sec:connection}, BCGs present minimal dynamical differences from cluster to cluster. On the contrary, as we move towards the outskirts, the contribution from the ICL increases. This component feels more the effects of the overall cluster potential, which is more sensitive to the global dynamical state of the cluster and the accretion pattern from the surrounding large-scale structure \citep[e.g.][]{evrard2008virial,saro2013toward}, thus producing a larger scatter in the velocity dispersion profiles. 
        Given that we are especially interested in examining the inner regions, we decided to stack the stellar profiles without any suitably rescaling of their amplitudes. Conversely, DM particles and galaxies velocity dispersion profiles require to be normalized to the cluster radius before stacking (provided that $\sigma(r) \propto r^{-1}$, as discussed in details in Appendix A of \citealt{marini2021phase}). Once we have the stacked median profiles of simulated clusters, we multiply the resulting profile by the virial radius of Abell S1063, \r200 = 2.36 Mpc, to properly compare to observational results. We see that in the internal regions, DM particles have a larger velocity dispersion than the stellar component, a consequence of the dissipative collapse which forms the BCG stars and determines their "colder" dynamics. As for substructures, since they undergo tidal effects during the merger process and they are disrupted when reaching the cluster center, the profile abruptly falls for $r/R_{200}< 0.1$ . Moving away from the cluster center, the three components converge to similar profiles. At large radii, the median profile spans values of the velocity dispersion which are lower than those in the observed cluster, similarly to what already encountered in Figure \ref{fig:Sartoris+2020}. Differences at this level can be again ascribed to several factors which may include distinct dynamical state and/or projection effects. At these scales, we find the substructures velocity dispersion profile to be consistently higher than that from the DM case. Analogously to what discussed in Section \ref{sec:connection}, we observe a velocity bias between the two tracers due to the effect of tidal stripping which is more effective in substructures with relatively low orbital velocity. The resulting mechanism is the selective removal of lower-velocity substructures, and thus an increase in the total velocity dispersion profile as traced by substructures.
        
    \subsection{Gradients of the stellar velocity dispersion profile}   
    
        \begin{figure*}
            \includegraphics[scale=0.5,angle=0.0]{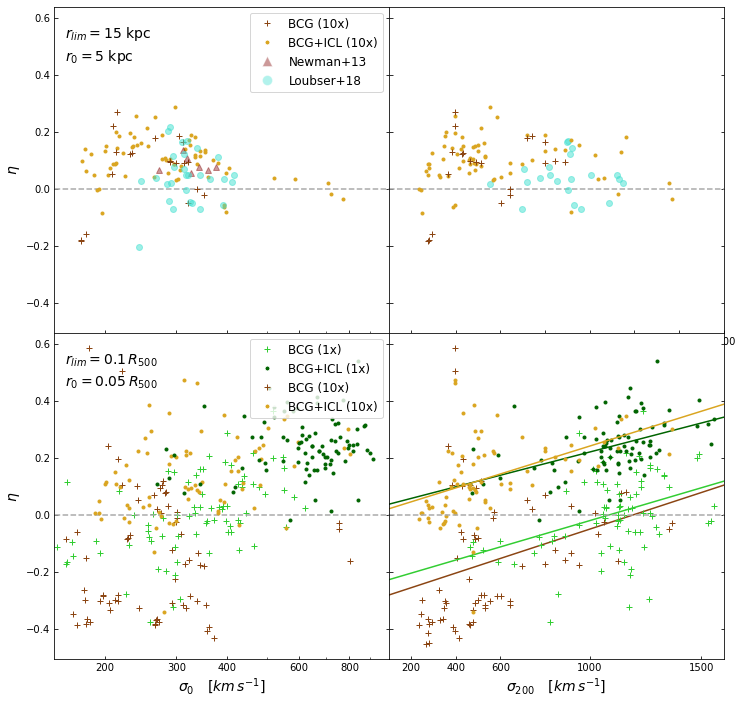}
            \caption{Top panels: comparison of the slope of the stellar velocity dispersion $\eta$ with the observational results from \protect\cite{newman2013density} (maroon triangles) and \protect\cite{loubser2018diversity} for BCGs (in light turquoise circles). The legend is common to all panels: dark green and golden points indicate the Hydro-1x and Hydro-10x's BCG+ICL samples, respectively; light green and brown crosses label the BCG-only sample from the Hydro-1x and Hydro10x runs, respectively. Following the approach adopted in \protect\cite{loubser2018diversity}, we report the slopes $\eta$ obtained by fitting the velocity dispersion profiles over the radial range from a central aperture of $r_0=5$ kpc and 15 kpc. We show results only for the Hydro-10x sample. Bottom panels: the same analysis as in the top panels, but computing the slope of the velocity dispersion profiles between $0.1\, R_{500}$ and $r_0=0.05\, R_{500}$, with the solid lines showing the best-fit results. The Left and right panels show the slope $\eta$ of the velocity dispersion profiles vs. the central velocity dispersion $\sigma_0$ and the cluster velocity dispersion \dispCL, respectively.}
            \label{fig:Loubser+2019GaussRov3}
        \end{figure*}

        Velocity dispersion profiles of the BCGs relate to the dynamical mass profiles of clusters and are expected also to be determined by the formation history of such extremely massive galaxies. Therefore, it is important to investigate and deepen the study on the velocity dispersion profile and its gradient. Up to now, studies of velocity dispersion profiles of BCGs exhibit a large variety of slopes, with a significantly larger fraction of positive slopes for BCGs, with respect to other early-type galaxies and brightest group galaxies (BGGs) \citep[e.g.][]{von2007special,bernardi2009evolution,huertas2013evolution}. 
        
        We present here the analysis of the slopes of the velocity dispersion profiles of BCGs in our set of "Hydro-10x" simulated clusters and compare with observational results obtained by \cite{loubser2018diversity}.
        Following these authors, we adopt the following model to describe the projected velocity dispersion profiles of BCGs:
        \begin{equation}
            \log \left(\frac{\sigma(r<r_{lim})}{\sigma_0} \right) =\eta \,\log  \left(\frac{r}{r_0} \right)\,.
        \end{equation} 
        Here the central velocity dispersion $\sigma_0$ is measured within an aperture $r_0$ from the BCG center. \cite{loubser2018diversity} (and \citealt{newman2013density} therein) obtained the BCG velocity dispersion within a radius $r_{lim}=15$ kpc and apply a central aperture $r_0=$ 5 kpc. This choice is motivated by previous studies \citep{graham1996brightest} which measured a typical half-light radius $R_e$ of 16.7 kpc in 119 Abell clusters.
        
        As for the analysis of simulations, we decided to follow two different approaches. In one case, for the comparison with observational results, we select a bi-dimensional radial aperture $r_{lim} = $ 15 kpc and compute the profiles on the Hydro-10x runs only. We exclude from this analysis the Hydro-1x clusters since their force resolution does not allow to adequately resolve the scales involved in the observational analyses. In the second case, we let $r_{lim}$ to vary with the size of cluster by taking $r_{lim}=0.1 R_{500}$: this value proved to include the BCG self-consistently with cluster size \citep[][and references therein]{ragone2018bcg}. In this case, the central aperture is also chosen to scale with the cluster radius, while being significantly larger than the softening length of both simulations with $r_0 = 0.05 R_{500}$. While this analysis can not be directly compared with observational results, it allows us to quantify the role of the ICL in determining the profiles over fixed fractions of the characteristic cluster scale radius.
        
        The velocity dispersion slope $\eta$ as a function of the central velocity dispersion $\sigma_0$ (or cluster velocity dispersion \dispCL) is reported in the top left (right) panel of Figure \ref{fig:Loubser+2019GaussRov3}. The legend is common to all panels: in golden the Hydro-10x, in dark green the BCG+ICL sample of the Hydro-1x and the BCG sample with light green crosses; maroon triangles and turquoise circles are the \cite{newman2013density} and \cite{loubser2018diversity} observational sets respectively. The top panels show the results of the analysis with $r_{lim} = $ 15 kpc, while the bottom panels refer to the analysis based on the different cut on the radial aperture $r_{lim}$. In the former, simulations predict slightly positive slopes, a result generally consistent with observational results from \cite{loubser2018diversity} and \cite{newman2013density}. We also note that our set of simulated BCGs shows few cases of negative slopes that often correspond to distinct projections of the same clusters having a three-dimensional negative radial gradient. These clusters do not seem to yield any distinct features with respect to the other clusters in the simulated set, except for the different gradients.

        In the bottom panels, we report the results from all runs (i.e. Hydro-1x and Hydro-10x, for both BCG+ICL and BCG cases), given that they should not be contaminated by resolution effects for the chosen radial range over which profiles are analyzed. On the left, we show the slope $\eta$ as a function of the central velocity dispersion $\sigma_0$: allowing $r_0$ to scale with $R_{500}$ yields a different sampling of $\sigma_0$ that extends to larger values than for the observational analysis shown in the upper panels. Furthermore, we observe that the BCG-only sample generally yields lower values of the gradients with respect to their counterpart BCG+ICL. Similar results are obtained in the right bottom panel which analyzes the relationship between $\eta$ the total cluster velocity dispersion \dispCL. The generally larger values of $\eta$ at fixed $\sigma_0$ (or $\sigma_{200}$) when ICL stars are included is due to the contribution of this dynamically warmer stellar component that increases the velocity dispersion when measured along the line-of-sight. At the same time, removing the ICL contribution still allows finding both positive and negative $\eta$, thus indicating that positive gradients are not necessarily due to the larger velocity dispersion of the ICL, but also reflect intrinsic dynamical properties of the BCGs. Additionally, here both Hydro-1x and Hydro-10x distributions hint at a positive correlation between $\eta$ and \dispCL, suggesting that more massive halos have increasing BCG velocity dispersion slopes.
        This correlation is confirmed by the Spearman coefficient $R_s$ computed on 100 bootstrapped samples of the simulated data which gives $R_s\simeq 0.5$ and p-value $\sim 0.02$. To help the eye, we report the best-fit lines for each data sample separately, following the color legend used with the points. A steeper slope in larger systems can be interpreted in light of Figure \ref{fig:Sohn+2019redshift_HIST}, which shows that BCG velocity dispersion increases less than linearly with the overall cluster velocity dispersion. This implies that in more massive clusters the relative contribution of the ICL to the observed $\sigma_\textrm{BCG}^{\star}$ increases. As a consequence, we expect this to translate into a larger contribution of the ICL to the projected velocity dispersion as we move toward larger projected radii, thus producing steeper slopes of $\sigma_\textrm{BCG}^{\star}(r)$. A similar argument is proposed by \cite{loubser2018diversity} when including a sample of brightest group galaxies in their analysis. These smaller galaxies -- measured within an aperture of 10 kpc and $r_0=$1 kpc which cannot be thoroughly compared with our simulations due to resolution limits -- exhibit a similar gradient. This is also in line with other observational studies, such as \cite{veale2017massive} who investigated the correlation between velocity dispersion profile slopes with galaxy environment and luminosity.

\section{Conclusions}
\label{sec:conclusions}

In this study, we address the reliability of simulating the physical properties of the Brightest Cluster Galaxies (BCGs), stressing the importance of the role played by the stellar velocity dispersion. The analysis is performed with a suite of cosmological hydrodynamical simulations, the DIANOGA set \citep[][]{ragone2018bcg,bassini2020dianoga}, obtained with the code GADGET-3 at two distinct resolutions (the base resolution is called Hydro-1x, the higher one is Hydro-10x). The astrophysical modelling includes radiative cooling, star formation and stellar feedback models \citep{springel2003cosmological}, metal enrichment and chemical evolution \citep{tornatore2007chemical}, and AGN feedback \citep{ragone2013brightest}. We are mostly interested in studying the stellar component associated with the main halo in the inner core ($< $ 50 kpc) of galaxy clusters which is composed of the stars bound to the BCG and the ICL, the diffuse stellar envelope that wraps the cluster halo. According to our definitions, we name \ll'BCG+ICL\rr' this stellar envelope, which is mostly consistent with what is detected in observations, while we apply an unbinding procedure to disentangle the two components, that are expected to yield distinct dynamics and formation histories, to only keep the stars bound to the BCG, referring to this sample as \ll'BCG\rr' or \ll'BCG-only\rr'. This separation between BCG and ICL stellar components is implemented with the purpose to examine the impact of excluding the warmer stellar component, i.e. the ICL, from the BCG velocity dispersion, a step that is extremely challenging in observational analyses.

The main results of our analysis can be summarized as in the following.
\begin{itemize}
\item Our DIANOGA set of simulated clusters has proven to reproduce fairly well the "cold" stellar dynamics of the BCG when considering the correlation between the BCG velocity dispersion and cluster velocity dispersion, \dispBCG -- \dispCL, as shown in Figure \ref{fig:Sohn+2019-gr0Gauss}: stellar velocity dispersions of simulated galaxy clusters agree with observational results from a set of local BCGs by \citet{sohn2020velocity}. We find this agreement to hold, even if simulations tend to produce too massive BCGs, when compared to observations (in Figure \ref{fig:scalings}).
\item  The \dispBCG-- $\sigma_{200}$ relation is independent of redshift and features a tight distribution with \dispBCG mostly spanning $\sim 100-600$ km s$^{-1}$, as reported in Figure \ref{fig:Sohn+2019redshift_HIST}. This is supported by the findings in Figure \ref{fig:dispBCG_z0-dispCL_z0} when following the late accretion of the most massive FoF groups in several Lagrangian regions from redshift $z=2$: they exhibit BCG velocity dispersions increasing at most by 30 percent, against the cluster velocity dispersions that, on average, double their values over the same redshift interval. The slow growth of the central galaxy since $z=$1 is only occasionally interrupted by late-time mergers that impact the value of BCG and/or cluster velocity dispersion. 
\item Our simulations describe quite well the observed dynamics yielded by distinct tracers, i.e. stars, DM particles, and galaxies. From Figure \ref{fig:Sartoris+2020} and \ref{fig:Sartoris+2020-stack} we argue that is especially true for stars in the innermost regions which make for most of the dynamically cold stellar component in the central galaxy. At larger radii, the presence of the diffuse ICL increases causing a rise in the velocity dispersion profiles of stars which, in turn, tend to recover a shape more similar to that of the velocity dispersion traces by DM and galaxies.
\item  Galaxy clusters, in both simulations and observations, are characterized by a large variety of profiles in the radial stellar velocity dispersion, as seen in Figure \ref{fig:Loubser+2019GaussRov3}. Considering the central stars (the stellar component within a bi-dimensional aperture of 50 kpc), we find a preference for positive gradients which is only partially due to the increasing velocity dispersion from the ICL, but rather it is an intrinsic property of the dynamics of BCGs. Furthermore, simulations reproduce the observed correlation between the slope of the velocity dispersion profile, $\eta$, and the cluster velocity dispersion (and thus, cluster mass) when the stellar velocity dispersion is computed in projection within a circular aperture whose size scales with cluster radius (i.e. $0.1R_{500}$).
\end{itemize}

One of the main conclusions of this work is that state-of-art simulations provide a reliable tool to interpret the dynamical processes operating at the center of galaxy clusters, and determining the projected phase-space structure of the BCG and the surrounding diffuse stellar component making up the ICL. These regions are particularly interesting for the evolutionary phenomena taking place (e.g. dynamical friction, mergers, galactic cannibalism) which largely impact cluster properties at these scales. To this end, studying the formation and evolution of the stellar content of the BCG (and surrounding ICL) can provide us insights into these mechanisms and their effect on the cluster as a whole. Our results show that stellar velocity dispersions robustly connect to the cluster velocity dispersion, as seen in observational analyses \citep[][]{loubser2018diversity}. Furthermore, recent measurements of stellar velocity dispersion profiles in the inner core of galaxy clusters, based on integral field spectroscopy \citep{sartoris2020clash}, have enabled us to directly compare high precision measurements with the simulated stellar population at these scales, yielding remarkably good results. 

Tracing the BCG and ICL dynamics in the innermost regions of galaxy clusters through detailed spectroscopic observations is expected to shed light on the processes driving star formation and leading to the assembly of the BCG since the infancy of proto-clusters at redshift $z>2$. In addition, the reconstruction of mass profiles at small cluster-centric radii should also provide information on the nature of DM, once the role of baryons is properly accounted for. In this respect, high-resolution cosmological hydrodynamical simulations provide an ideal tool to follow the process of BCG assembly and the interplay between DM and stellar dynamics. As highlighted by the analysis presented in this paper, such simulations can account for many dynamical properties of the stellar component inside and around BCGs. On the other hand, a general limitation of the current generation of simulations, including ours, is that they tend to overpredict BCG stellar masses. This calls for the need to further improve the numerical description of feedback mechanisms, most probably related to AGN, included in such simulations for them to keep the pace of the increasing quality and quantity of data expected from the next generation of observational facilities. 

\section*{Acknowledgements}
We would like to thank Luigi Bassini, Andrea Biviano, and Veronica Biffi for useful discussions. 
This project has received funding from the European Union's Horizon 2020 Research and Innovation Programme under the Marie Sklodowska-Curie grant agreement No 734374. We acknowledge financial support from the INFN INDARK grant, PRIN-MIUR 2015W7KAWC, the EU H2020 Research and Innovation Programme under the EuroEXA project (Grant Agreement ID: 754337), the Italy-Germany MIUR-DAAD bilateral grant n. 57396842. AS is supported by ERC-StG ‘ClustersXCosmo’ grant agreement 716762, by the FARE-MIUR grant 'ClustersXEuclid' R165SBKTMA. KD acknowledges support by the Deutsche Forschungsgemeinschaft (DFG, GermanResearch Foundation) under Germany's Excellence Strategy -- EXC-2094 -- 390783311 as well as for the COMPLEX project from the European Research Council (ERC) under the European Union’s Horizon 2020 research and innovation program grant agreement ERC-2019-AdG 860744. CRF and GLG thank the Consejo Nacional de Investigaciones Cient\'ificas y T\'ecnicas de la Rep\'ublica Argentina (CONICET) for financial support. Y.W. is supported by NSFC grant No.11803095, NSFC grant No.11733010.
Simulations have been carried out using MENDIETA Cluster from CCAD-UNC, which is part of SNCAD-MinCyT (Argentina); at CINECA (Italy), with CPU time assigned through ISCRA-B grants, and through a University of Trieste-CINECA agreement; at the Tianhe-2 platform of the Guangzhou Supercomputer Center by the support from the National Key Program for Science and Technology Research and Development (2017YFB0203300). We acknowledge CINECA and INAF, under the coordination of the "Accordo Quadro MoU per lo svolgimento di attività congiunta di ricerca Nuove frontiere in Astrofisica: HPC e Data Exploration di nuova generazione", for the availability of computing resources and support, and the project INA17\_C5B32. We acknowledge the computing center of INAF-Osservatorio Astronomico di Trieste, under the coordination of the CHIPP project \citep{bertocco2019,taffoni20}, for the availability of computing resources and support.  

\section*{Data Availability}

The data underlying this article will be shared on reasonable request to the corresponding author.



\bibliographystyle{mnras}
\bibliography{Bibliography} 




\appendix
\section{Machine learning technique to identify BCG and ICL}
\label{AppendixB}

In simulations, one can exploit the full dynamical information available on star particles to separate the two stellar populations belonging to the BCG and the diffuse ICL. An example of such methods is given by the modified version of the halo finder Subfind (as described in \citealt{dolag2010dynamical} and Section \ref{subsec:stellarsubfind} in this work). The algorithm identifies the single star particles in the main halo of clusters as either bound to the BCG or ICL, by applying a dynamical criterion. The underlying assumption is that the two velocity distributions of stars belonging to the ICL and the BCG have each a Maxwellian shape so that the overall velocity distribution of stars is described by a double Maxwellian. The diffuse ICL is associated with the Maxwellian with the larger velocity dispersion, in contrast, the BCG, having colder dynamics, populates the distribution at lower dispersion. To assign each stellar particle to either one of the two dynamical components, the algorithm follows an unbinding procedure, by iteratively computing the gravitational potential given by all particles within a certain radius. The latter identifies two stellar populations, classified as "bound" and "unbound", which are separately described by a single Maxwellian. If the best-fit parameters of the initial double Maxwellian coincide with those obtained in the single Maxwellians, then the procedure is completed, otherwise, the radius of the sphere is changed accordingly and the computation is remade. 

Given the nature of this algorithm, based on the well-defined properties of the stellar components, it can be also replicated by a machine learning (ML) model in a faster and more efficient way. The goal is to provide an alternative classification method to identify stars in the main halo according to these same features. To achieve this, we design a supervised Random Forest Classifier\footnote{from the scikit-learn package \citep{scikit-learn} } having as input features the cluster mass \m200, the cluster-centric spatial distance and velocity distribution of each particle, and as output the binary label BCG or ICL. This combination of features fed to the machine is the one that provides the best and more consistent results with respect to the true label obtained by Subfind. More details on this ML-based algorithm can be found in Marini et al. (2021, in preparation), while here we provide only a short description for completeness of information.

\subsection{Data and training}
To train the algorithm we use the data from the clusters set already presented in this work with the name "Hydro-1x" (see Section \ref{sec:simulations}). We randomly select a subset of the star particles from five of the 29 clusters and we feed the model with the input features (\m200, cluster-centric distance, and velocity of each particle) from the stars in the main halo. The cluster-centric distance and velocity are both normalized to exclude effects due to the different sizes of the clusters in the training set. The former is scaled for \r200, while the velocity is normalized for $V_{200}=(G M_{200}/R_{200})^{1/2}$, which is the circular velocity of the cluster at \r200. Then, we test the results on the remaining clusters, reaching for each cluster an accuracy always in the range of 80-85 percent. Additionally, we check that the initial choice of which cluster to use for training does not affect the resulting metric of the model (changes in accuracy are up to a maximum of $\sim$ 10 percent). 

\subsection{Results}
Although this appendix aims at providing only the key aspects of this model, we show here a few examples of its performances, in comparison to those obtained when applying the true label (i.e. the Subfind algorithm). This shall clarify the resolving power of the model, compared to the traditional method. With this in mind, we randomly select one cluster from the Hydro-1x simulation and in the following, we will examine the comparison with feature distributions, density profiles, and density maps. 

\subsection{Cluster specific}
This cluster has mass $M_{200}=$ 1.43$\times 10^{15}$ \msun, orbital velocity at \r200 of $V_{200}=$ 1832 km s$^{-1}$, and its dynamical state is classified as intermediate, following the prescription provided in \cite{biffi2016nature}. In short, the method is based on two properties: a low center shift (the distance between the position of the minimum of the gravitational potential and the center of mass is lower than a chosen threshold value, in our case 0.07 \r200) and low fraction of mass in substructures ($M_\mathrm{TOT,sub}/M_\mathrm{TOT}<0.1$). A halo is classified as relaxed if both conditions are satisfied, intermediate if only one holds, and disturbed if none is satisfied. 
The ICL mass fraction estimated when using the stellar Subfind is 0.732, corresponding to the particle velocity distribution with a double Maxwellian best-fitted by \dispBCG = 575 km s$^{-1}$ and $\sigma_\mathrm{ICL}^{\star}$ = 1101 km s$^{-1}$. 

\subsubsection{Features distribution}
\begin{figure*}
\centering
    \includegraphics[scale=0.45,angle=0.0]{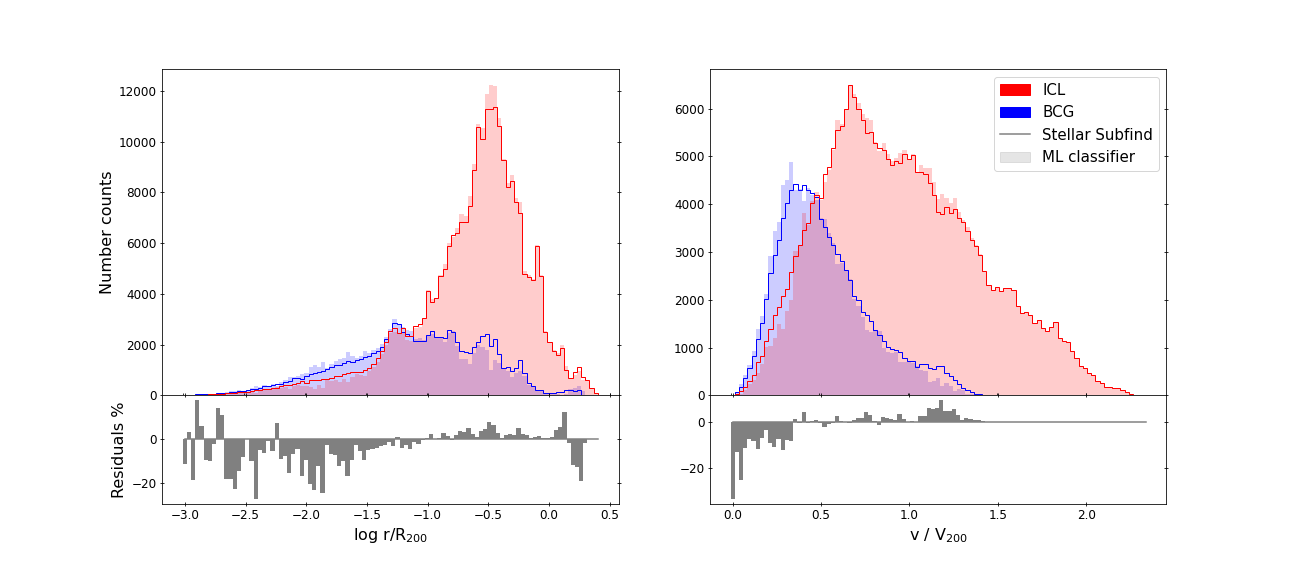}
    \caption{\emph{Top:} Distribution of the input features drawn from the Hydro-1x simulation. \emph{Bottom:} Percent residuals measured between the true and predicted counts of ICL particles in each bin. In each panel, we plot the predicted (bars) and true (line) number counts of both ICL (red) and BCG (blue) associated with the star particles. More in detail we have on the left the distributions of the logarithmic cluster-centric distance normalized by \r200 and on the right the stellar particle velocities normalized by $V_{200}$.}
    \label{fig:D14_features-1x}
\end{figure*}

Figure \ref{fig:D14_features-1x} shows the comparison between the distribution of cluster-centric distances (left panel) and velocities (right panel) of the star particles drawn from the histograms of the input features computed for the true (contouring line) and predicted labels (blocks) of both ICL (red) and BCG (blue). Additionally, we compute the residuals between true and predicted labels in the ICL over the total number of star particles in a given bin to estimate where the two distributions differ. 

In both cases, we observe a good agreement within each predicted subgroup and its true distribution. The left panel confirms the presence of a bulk structure in the inner regions, which corresponds to the BCG, and a larger more diffuse component extending past \r200, the ICL. The largest differences are identified in the inner core of the BCG (around 30 percent) but they are mostly due to the low number of particles in these bins. As for the velocity distribution, we confirm the presence of the two peaks which can be reproduced by a double-Maxwellian distribution. In this case, we see that the machine learning algorithm shows a preference for a larger separation between the two Maxwellian distributions, thus increasing the distance in the phase-space between the two stellar components. Yet, the measured difference in the residuals is never larger than 30 percent. 

Fitting a double Maxwellian in the velocity distribution of the particles with the predicted labels yields \dispBCG = 552 km s$^{-1}$ and $\sigma_\mathrm{ICL}^{\star}$ = 1112 km s$^{-1}$, values that are comparable to those in the true-label distribution. Analogously, the ICL mass fraction (= 0.718) is comparable to that in the true label case. 

\subsubsection{Density maps and profiles}
\begin{figure*}
        \includegraphics[scale=0.5,angle=0.0]{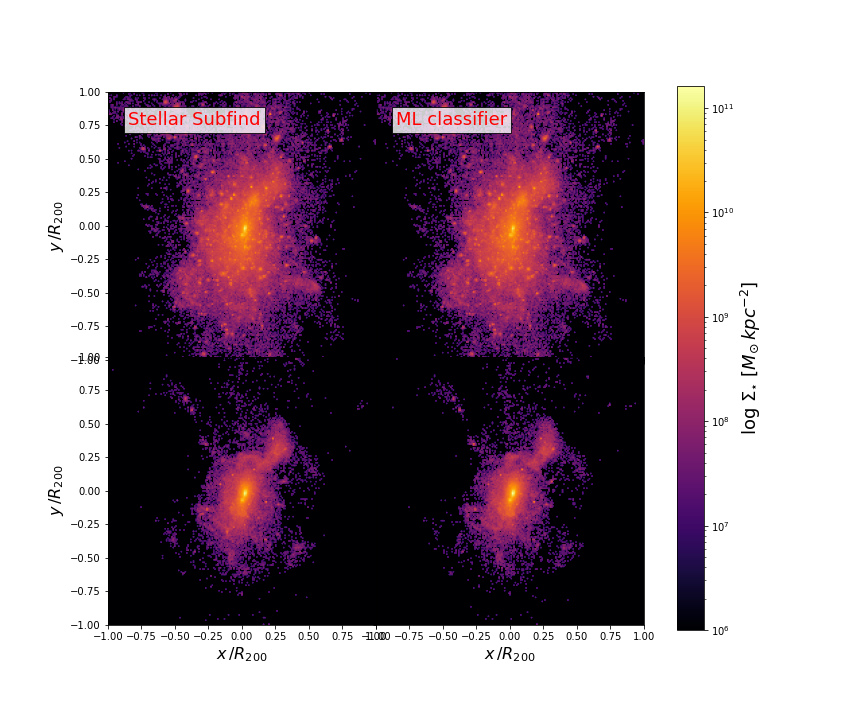}
        \caption{Density maps of the stellar components (ICL in the top panels, BCG in the bottom panels) in the cluster selected within the Hydro-1x simulation. The left panels report the results from the stellar division provided by the traditional method with the stellar Subfind. The right panels show the density maps for the stellar components identified with the predicted labels.}
        \label{fig:D14_map-1x}
    \end{figure*}

\begin{figure}
        \includegraphics[scale=0.5,angle=0.0]{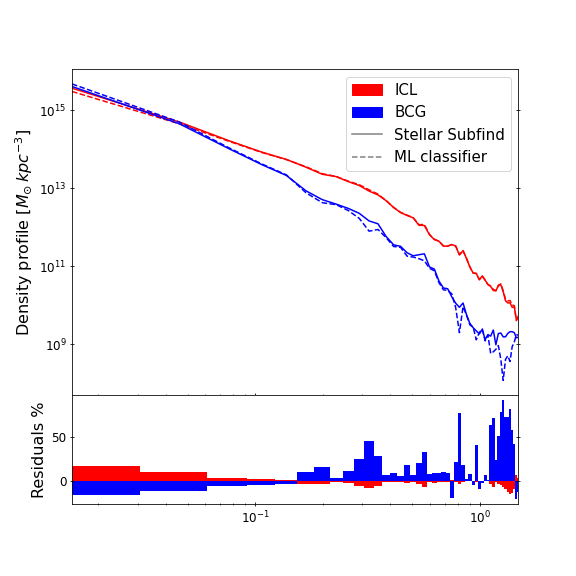}
        \caption{Density profiles of the BCG (blue) and ICL (red) in the cluster from the Hydro-1x simulation. The dashed lines are the profiles computed with the predicted labels, while solid lines report the profiles computed with the true labels. }
        \label{fig:D14_density-1x}
    \end{figure}

Figure \ref{fig:D14_map-1x} shows the density maps of the distinct stellar components in the cluster under study: ICL in the top panels and BCG in the bottom. The panels on the left are the results from the traditional Subfind method, while on the right we show the results from the machine learning algorithm. The results are remarkably similar and differences can be hardly detected unless one examines the details of the edges of the BCG distribution. This is more evident when comparing the three-dimensional density profiles of both ICL (red) and BCG (blue) as computed with the true labeled stars (solid line) or with the predicted labels (dashed line) in Figure \ref{fig:D14_density-1x}. We remind that star particles responsible for this difference are mostly located in the outskirt of the main halo, thus the difference is negligible.

The fraction of stars associated with the ICL by the ML method yields values that are also consistent with those predicted by Subfind: in this particular cluster we measure 0.758 and 0.745 for the ICL mass fraction predicted by the ML and the Subfind algorithm, respectively. Once the classifier is trained, we measure the CPU time to estimate the gain by employing this method. We run the classifier on 128 processors, accounting for a total CPU time of 52.8s employed to classify 366037 star particles in the main halo of the cluster.

\bsp	
\label{lastpage}
\end{document}